\documentclass[10pt,twocolumn,letterpaper]{article}

\usepackage{cvis}              
\usepackage{float}
\definecolor{cvisblue}{rgb}{0.21,0.49,0.74}
\usepackage[pagebackref,breaklinks,colorlinks,allcolors=cvisblue]{hyperref}
\def\confName{CVIS}
\def\confYear{2026}

\title{A Physics-Informed Digital Twin Framework for Calibrated Sim-to-Real FMCW Radar Occupancy Estimation}

\author{Huy Trinh\\
University of Waterloo\\
{\tt\small h3trinh@uwaterloo.ca}
\and
Sebastian Ratto V\\
University of Waterloo\\
{\tt\small srattovalderrama@uwaterloo.ca}
\and
Elliot Creager\\
University of Waterloo\\
{\tt\small creager@uwaterloo.ca}
\and
George Shaker\\
University of Waterloo\\
{\tt\small gshaker@uwaterloo.ca}
}

\begin{document}
\maketitle
\begin{abstract}
Learning robust radar perception models directly from real measurements is costly due to the need for controlled experiments, repeated calibration, and extensive annotation. 
This paper proposes a lightweight simulation-to-real (sim2real) framework that can enable reliable Frequency Modulated Continuous Wave (FMCW) radar occupancy detection and people counting using only a physics-informed geometric simulator plus a small unlabeled real calibration set. A calibrated domain randomization (CDR) step is introduced that aligns the global noise-floor statistics of simulated range--Doppler (RD) maps to those observed in real environments, while preserving discriminative micro-Doppler structure. 
Across real-world evaluations, ResNet18 models trained purely on CDR-adjusted simulation achieve 97\% accuracy for occupancy detection and 72\% accuracy for people counting, outperforming ray tracing baseline simulation and conventional random domain-randomization baselines. 
\end{abstract}











\section{Introduction}
\label{sec:intro}
\begin{figure*}[!t] 
  \centering    \includegraphics[
    width=\textwidth,
    height=.3\textheight,   
    keepaspectratio,         
  ]{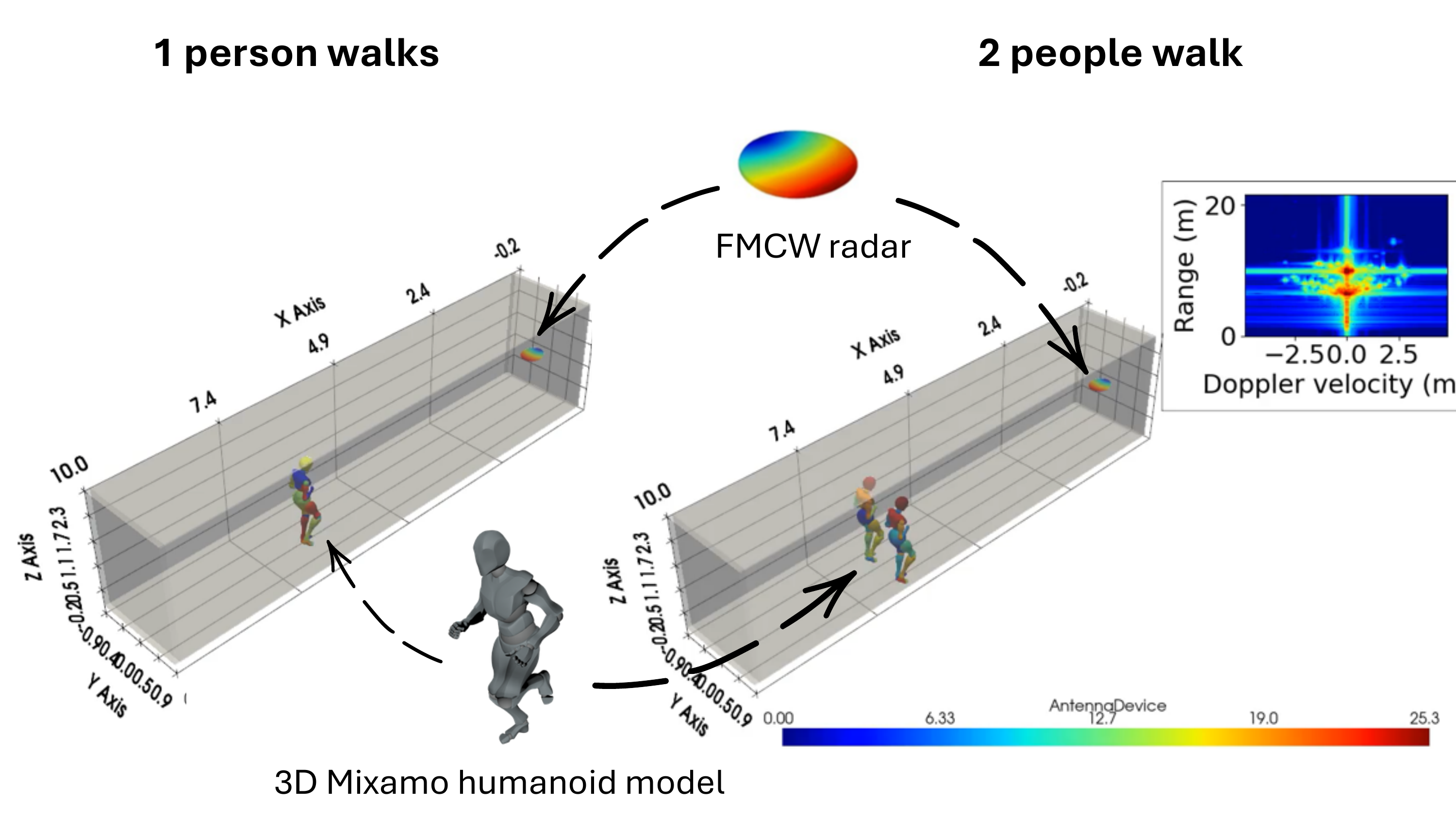}
  \caption{Simulation setup showing the created radar scenes. A 3D Mixamo humanoid model is animated to represent walking motion and imported into the virtual corridor environment. The FMCW radar observes different scenarios—(left) a single walking subject and (right) two subjects walking at different speeds, producing corresponding range–Doppler signatures.} 
  \label{fig:Digital_twin_plot}
\end{figure*}
High-fidelity simulators and radar digital twins based on full-wave or ray-tracing electromagnetic solvers promise scalable synthetic data generation in various perception applications. Given a geometric and electromagnetic model of the scene, a simulator can synthesize large quantities of labeled radar data under controlled conditions, including difficult or safety-critical scenarios~\cite{10506223, 10.1007/978-3-031-97772-5_3}. Recent works have leveraged full-physics simulation~\cite{Chipengo2018FromAD,9446140}, motion-capture driven micro-Doppler synthesis~\cite{10365512}, or Generative Adversarial Network (GAN) based generation~\cite{Erol2019GANbasedSR,9944857}. In addition, various studies focus on bridging the gap between simulation and reality on object detection, gesture, and human activity recognition~\cite{Bialer2024RadSimRealBT, Kern2022LearningOM, ElHail2025RadarBasedHA}. 
These works often depend on proprietary electromagnetic (EM) toolchains, extensive parameter sweeps, and specific domain knowledge tuning. This paper proposes a one-shot, calibration-based noise-floor alignment between simulation and reality to enable efficient sim-to-real transfer for FMCW radar occupancy detection and people counting. To support this, we employ a modular physics-informed digital twin framework that integrates human animation, electromagnetic modelling, and automated radar data generation. 
The remainder of this paper is organized as follows. Section~\ref{sec:methodology} details the digital twin simulated radar generation, real data collection, calibration formulation, and learning setup. Section~\ref{sec:results} presents results, and Section~\ref{sec:conclusion_future_work} concludes and discusses future research.

\section{Methodology}
\label{sec:methodology}


\subsection{Physics Informed Digital Twin for Radar Data Generation}

A physics-informed digital twin of the monitored corridor was developed using an electromagnetic simulation platform that combines Shooting and Bouncing Rays (SBR) and Physical Optics (PO) methods~\cite{11114146,11186561}. The corridor environment, including walls, ceiling, and floor, was reconstructed with measured dielectric properties to capture realistic multipath propagation and clutter effects. A virtual 60~GHz FMCW radar modeled after the Infineon BGT60TR13C sensor~\cite{InfineonBGT60TR13C} replicated the antenna configuration, polarization, and chirp parameters of the physical hardware.
The radar was emulated with a carrier frequency of 60~GHz and bandwidth of 882.35~MHz. The sampling rate and chirp duration were set to 1~MHz and 0.256~ms, respectively, yielding 256 samples per chirp and a chirp slope of approximately $3.44 \times 10^{12}$~Hz/s. These parameters correspond to a theoretical range resolution of 0.17~m and a maximum unambiguous range of 21.7~m, consistent with the operating characteristics of the real device.
Human motion dynamics were introduced using animated 3D humanoid meshes derived from Mixamo and CMU Motion Capture datasets~\cite{Mixamo,CMU_Mocap}. These animated models provided time-varying radar cross sections and realistic trajectories for walking and interaction behaviours. The electromagnetic solver iteratively computed scattered fields through multiple ray interactions until the residual reflected power dropped below a predefined threshold.
The simulated radar outputs were organized as complex data cubes containing amplitude and phase information across fast time, slow time, and frequency dimensions. Range Doppler maps were then generated for representative scenes, including static clutter, a single walking subject, and two subjects moving toward the radar, as Fig.~\ref{fig:Digital_twin_plot}. All simulated outputs were exported as floating-point arrays for subsequent machine learning analysis and calibration with real radar measurements. In the remainder of the paper, we refer to this method as \textbf{ray-tracing baseline (RT baseline)} or \textbf{RT simulation}.

\subsection{Real Data Collection and Processing Pipelines}

\begin{figure}[htb]
    \centering
    \includegraphics[scale=0.35, height=.22\textheight, width=0.25\textwidth]{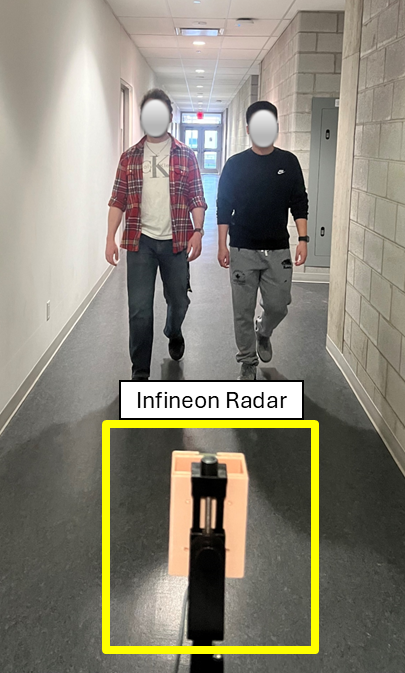}
    \caption{Experiment setup for 2 people walking scenario.}
    \label{fig:real_2_people_walk_corridor}
\end{figure}

\begin{figure*}[!t]
  \centering    \includegraphics[width=0.8\textwidth, height=0.21\textheight 
  ]{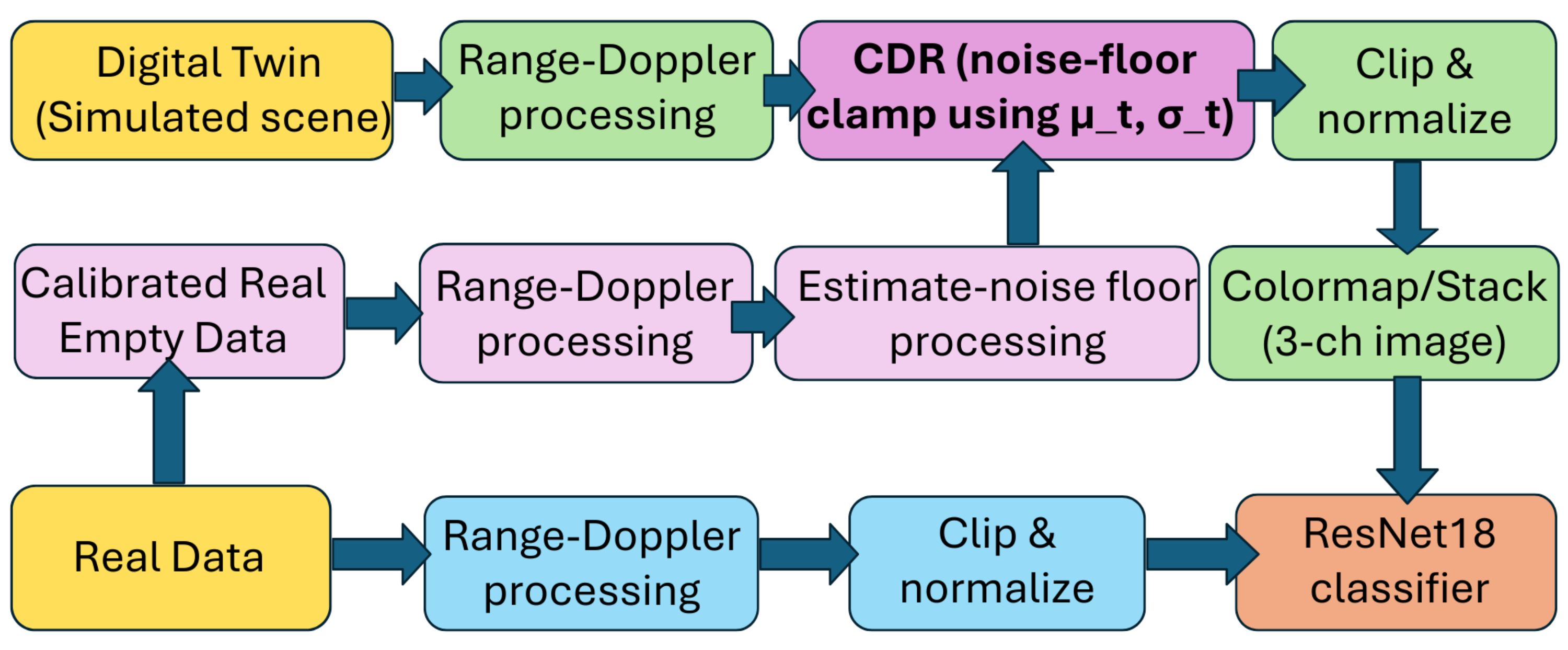}
    \caption{Overview of the proposed sim-to-real pipeline. \textbf{Yellow blocks:} Start of simulated data and real data preprocessing \textbf{Green path}: simulated radar frames from the digital twin are converted to range–Doppler (RD) maps and passed through the noise-floor CDR block, then clipped, normalized, and colour-mapped for network training. \textbf{Purple path}: CDR uses 10 seconds of unlabeled real empty-room data to estimate the target noise-floor statistics that drive the calibration. \textbf{Blue path}: real radar frames for evaluation are converted to RD maps and only clipped and normalized, without any randomization or CDR modification.}
  \label{fig:CDR_flowchart_2}
\end{figure*}

A real-world FMCW radar dataset is collected in the corridor using the same Infineon BGT60TR13C radar~\cite{InfineonBGT60TR13C} and configuration specified above. We record three occupancy states: empty room, one person walking, and two people walking the radar sensor as seen in Fig.~\ref{fig:real_2_people_walk_corridor}.
An additional short unlabeled calibration sequence (10 seconds or 100 frames) of an empty corridor is acquired once and used only to estimate the real noise-floor statistics for CDR (Sec.~\ref{subsec:CDR}).
Both simulated and real data are applied with a 2D FFT applied across fast-time and slow-time dimensions to obtain a complex Range-Doppler map, and their magnitudes are converted to dB. The real data processing stage does not go through any randomization and is used exclusively for testing purposes. They are then clipped to a fixed dB range $[v_{\min}, v_{\max}]$ estimated from simulation, normalized to $[0,1]$, and mapped via a perceptually uniform colormap (viridis) to form 3-channel images for the input of the neural network model (ResNet18 with random initialization and a task-specific classifier head). 
The final fully connected layer is replaced with a task-specific classifier:
2 units for binary occupancy or 3 units for people counting applications. Training is performed on simulated RD maps: either (i) RT simulation, (ii) RT simulation with Random Domain Randomization (Random DR), or (iii) RT simulation with the proposed Calibrated Domain Randomization (CDR).
The real labelled test set is held out entirely for sim-to-real evaluation. Real RD maps at test time are passed through the same clip-and-normalize pipeline as the simulated inputs (Fig.~\ref{fig:CDR_flowchart_2}).
\subsection{Noise-Floor Domain Randomization and CDR}
\label{subsec:CDR}
\begin{figure}[htb]
  \centering    \includegraphics[scale=0.354]{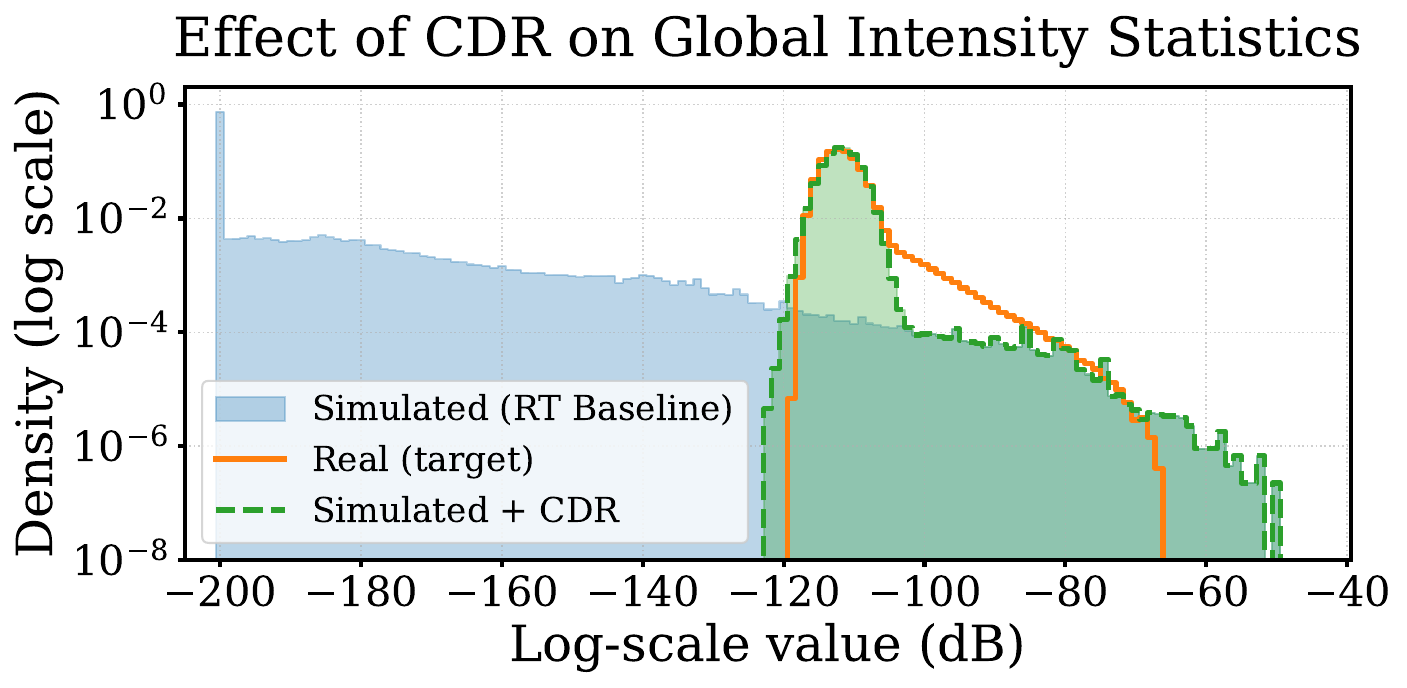}
  \caption{Effect of CDR on global RD magnitude statistics. Baseline simulation (blue) has quite a different low noise floor. CDR (green) shifts the simulated histogram toward the real empty-room distribution (orange).}
  \label{fig:hist_cdr_effect}
\end{figure}
\begin{figure}[htb]
  \centering    \includegraphics[scale=0.4,
  width=0.45\textwidth]{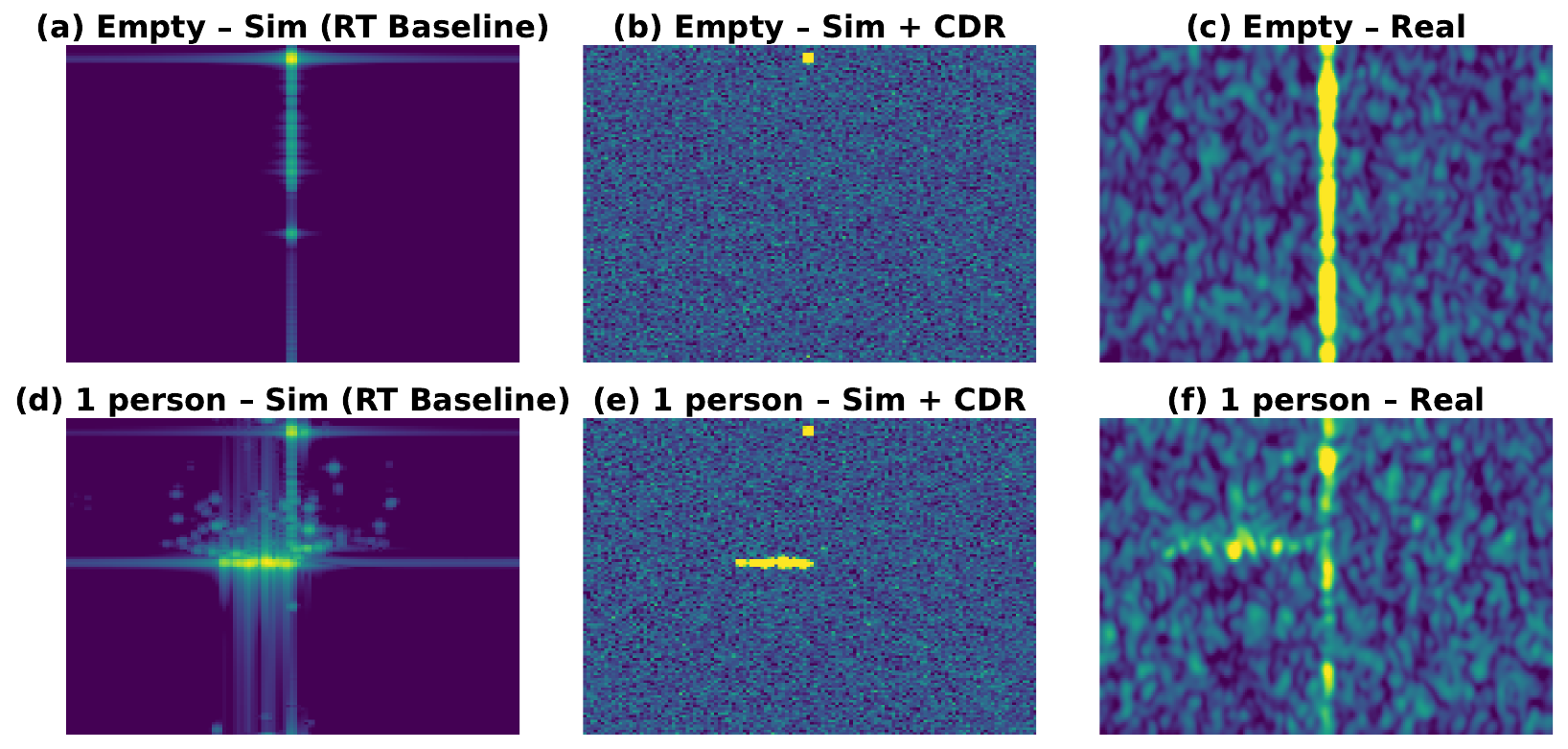}
  \caption{Range-Doppler maps for (a,d) baseline simulation, (b,e) simulation
    after CDR, and (c,f) real measurements for empty room and one-person
    walking.}
  \label{fig:sim_cdr_real_triplets}
\end{figure}
\begin{figure}[htb]
    \centering
    \includegraphics[scale=0.3]{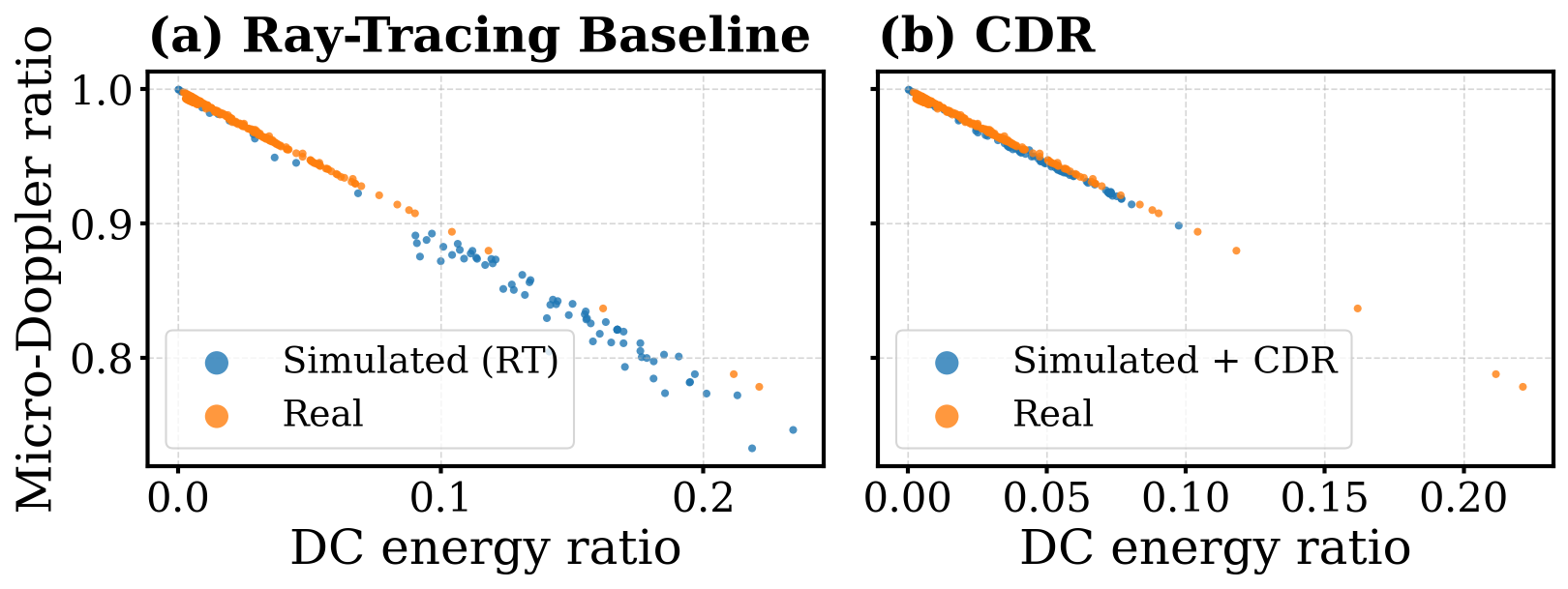}
    \caption{Static-to-total energy ratio in RD space. Each point is one frame. CDR moves simulated points toward the real cluster, indicating that the global clutter level has been calibrated.}
    \label{fig:dc_energy}
\end{figure}
\begin{figure}[H]
    \centering
    \includegraphics[width=0.47\textwidth, height=0.18\textheight]{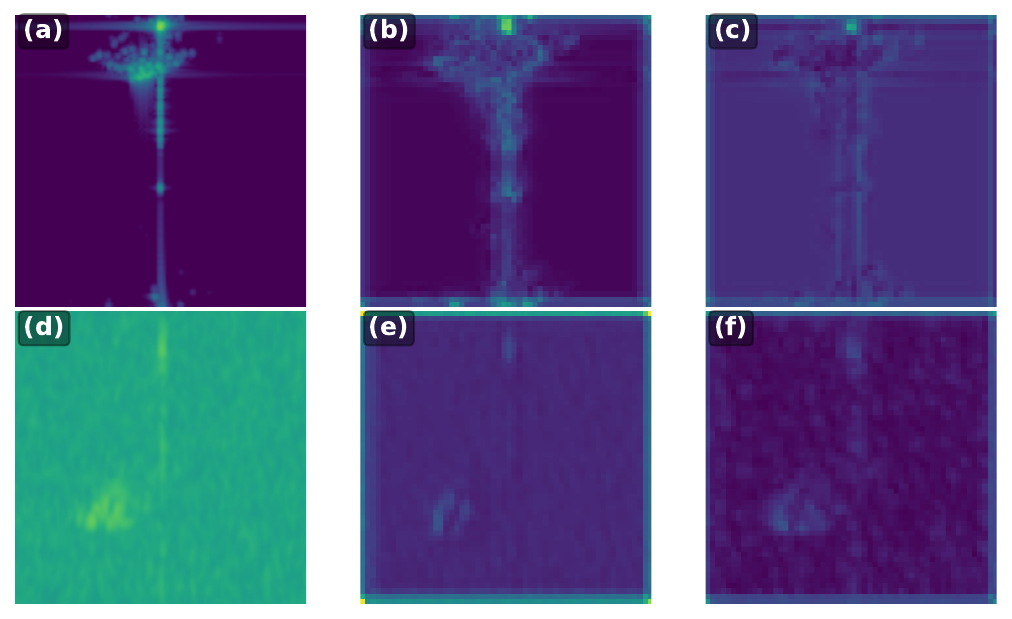}
    \caption{
    Qualitative visualization of early feature maps from the first residual block of ResNet-18 for representative simulated and real radar range-Doppler inputs.
    (a,d) inputs;
    (b,e) features from the baseline RT model trained only on unmodified simulation;
    (c,f) features from the CDR-trained model.}
    \label{fig:feature_maps}
\end{figure}

Inspired by domain randomization (DR) in computer vision and robotics \cite{Tobin2017DomainRF, DBLP:conf/rss/SadeghiL17,Tremblay2018TrainingDN}, our approach introduces synthetic noise to bridge this gap.
Both our baseline Random DR and a proposed Calibrated DR (CDR) method apply a per-frame, per-cell clamp in the dB domain, before clipping and normalization:
\begin{equation}
    \tilde{S}(i,j) = \max\big(S(i,j),\,N(i,j)\big).
    \label{eq:clamp_generic}
\end{equation}
This operation raises the overly clean simulated background cells $S(i,j)$ to a sampled noise-floor level $N(i,j)$ with range--Doppler cell $(i,j)$.
The two methods differ only in how they sample the noise map $N(i,j)$. For the Random DR baseline, we emulate generic domain randomization by sampling: 
\begin{equation}
    N(i,j) \sim \mathcal{N}(m, s^2)
\end{equation}
with mean $m$ from $[-130,-100]$\,dB and a standard deviation $s$ from $[1,4]$\,dB. These ranges approximate typical background levels and variability observed in our quantitative indoor FMCW experiment.
CDR keeps the same clamping form \eqref{eq:clamp_generic} but replaces
heuristic sampling with a one-shot, data-driven calibration from the unlabeled real empty-room calibration set by aggregating all RD magnitudes and computing their empirical
mean $m_t$ and standard deviation $s_t$. 
CDR then samples
\begin{equation}
    N_t(i,j) \sim \mathcal{N}(m_t, s_t^2)
\end{equation}
i.i.d.\ over $(i,j)$ and applies the same clamp
$\tilde{S}(i,j) = \max(S(i,j), N_t(i,j))$. Fig.~\ref{fig:hist_cdr_effect} shows that CDR shifts the simulated
histogram toward the real empty-room distribution, while baseline
simulation underestimates the noise floor. Fig.~\ref{fig:sim_cdr_real_triplets} further illustrates how CDR transforms
overly clean simulated RD maps into clutter patterns that visually match
real measurements while maintaining Doppler walking signatures. Fig.~\ref{fig:dc_energy} confirms that CDR moves the ratio of low-Doppler (clutter) energy to total energy for simulated samples closer to real data. Finally, to better understand how CDR alters the learned representation,
Fig.~\ref{fig:feature_maps} visualizes channel-averaged activations from the first residual block of ResNet-18 for representative simulated and real RD inputs. When applied to real measurements, the baseline model produces diffuse responses dominated by corridor clutter and simulator-specific artifacts.
In contrast, the CDR-trained model concentrates its responses on the range--Doppler cells associated with human motion while suppressing background structure in both domains.

\section{Results}
\label{sec:results}

\begin{figure}[t]
    \centering
    \includegraphics[scale=0.24]{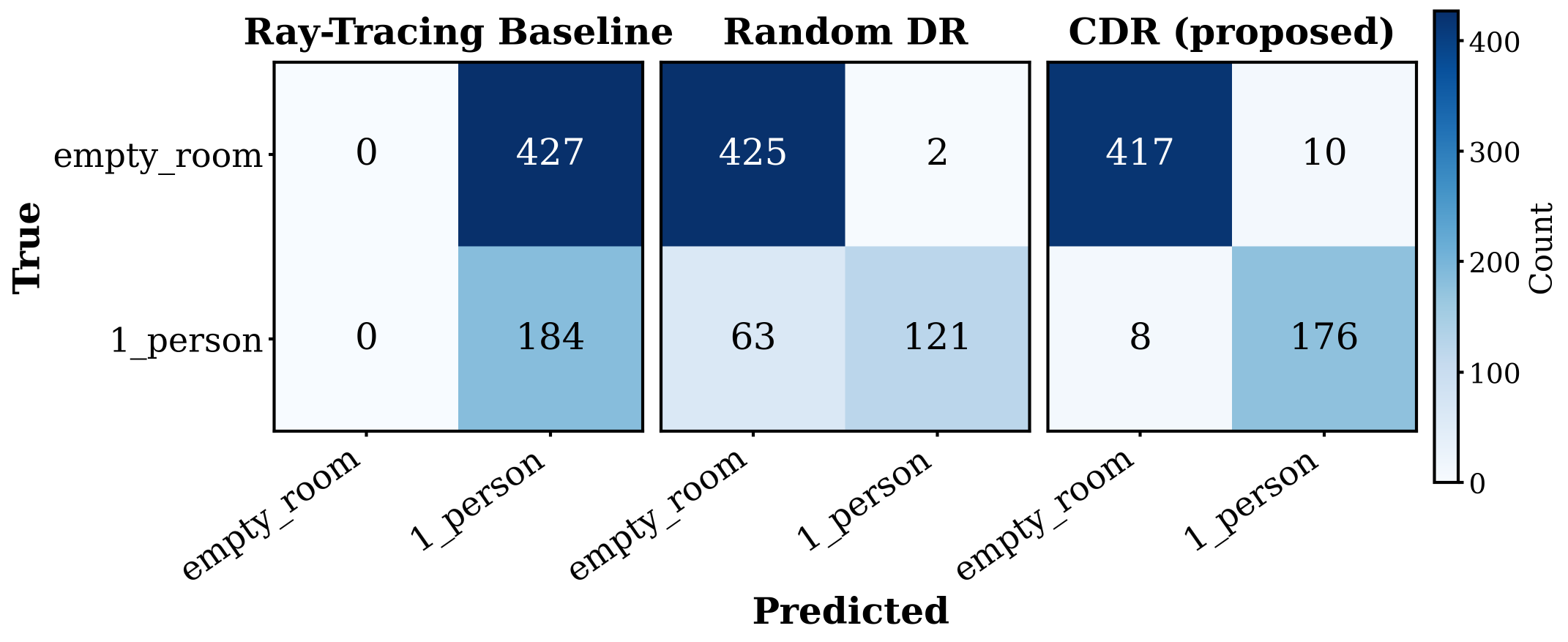}
    \caption{Confusion matrices for real binary occupancy detection across 3 methods.}
    \label{fig:conf_binary}
\end{figure}

On the real corridor test set, the baseline RT model trained solely on unmodified simulation struggles to generalize.
Fig.~\ref{fig:conf_binary} shows that it misclassifies many occupied frames as empty due to mismatched clutter and noise, resulting in a balanced accuracy close to 50\% (chance-level for a binary task). The introduction of uncalibrated Random DR improves robustness: the model becomes less sensitive to background mismatch and achieves balanced accuracy around 83\%. However, the resulting RD samples sometimes exhibit unrealistic structure, and performance is still limited by residual sim-to-real discrepancies. CDR provides a further and substantial improvement. By aligning simulated noise floors to real statistics while preserving micro-Doppler content, the CDR-trained model achieves balanced accuracy $\approx 97\%$ and high overall accuracy. The confusion matrices (Fig.~\ref{fig:conf_binary}) show that CDR dramatically reduces both false negatives and false positives compared to the baselines.
\begin{figure}[t]
    \centering
    \includegraphics[scale=0.26]{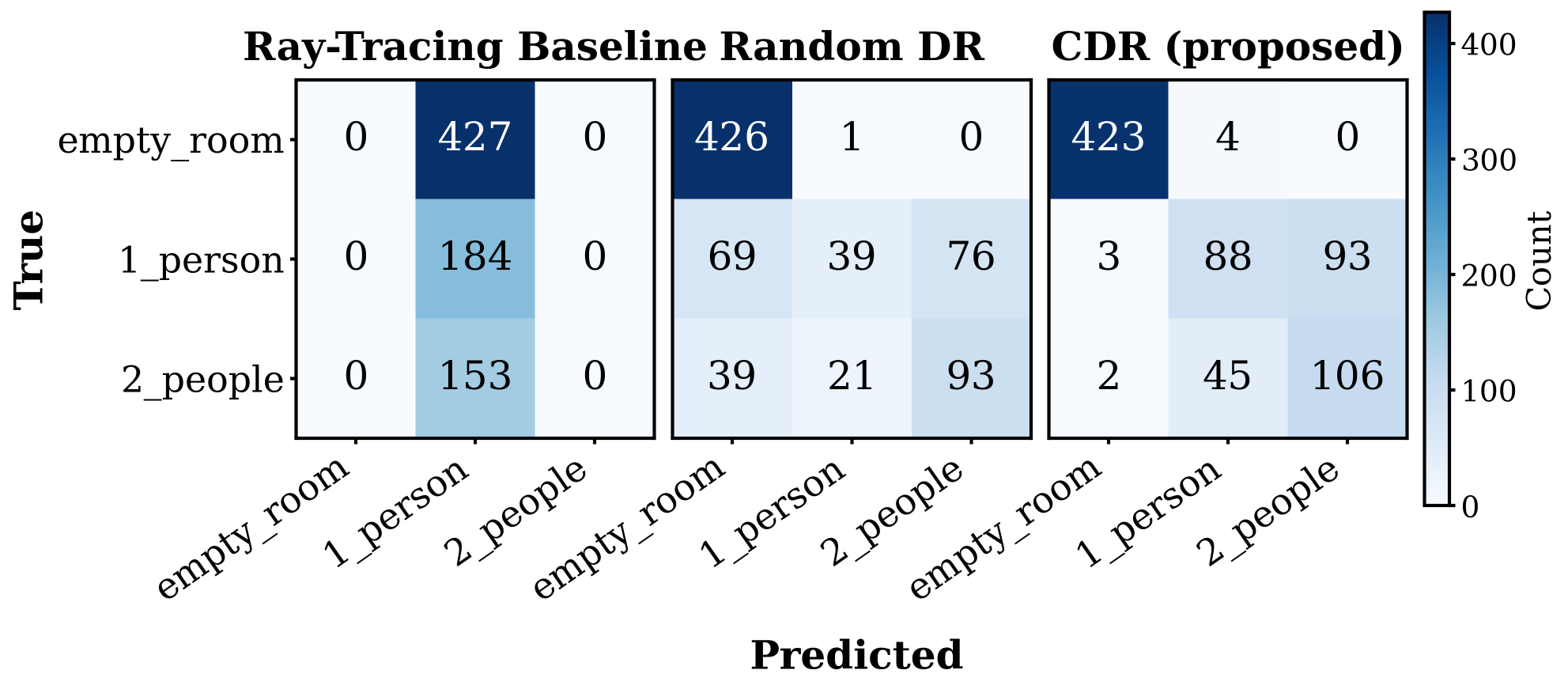}
    \caption{Confusion matrices for real people counting across 3 methods.}
    \label{fig:conf_count}
\end{figure}
The people counting task is more challenging due to the overlap between one- and two-person signatures and limited data.
Simulated training remains feasible by using same pipeline, while the real test set is used exclusively for evaluation. As shown in Fig.~\ref{fig:conf_count}, the baseline RT model again fails to transfer, achieving balanced accuracy near 33\% (random guessing) 
Random DR improves performance to balanced accuracy $\approx 61\%$, indicating that generic noise-floor perturbation offers some regularization but does not fully align domains, and heavily confuses among 3 classes. CDR yields the best performance with balanced accuracy $\approx 72\%$ and improved confusion matrices, particularly for distinguishing one vs.\ two-person cases. 
Misclassifications mainly occur in ambiguous scenes with partial overlap or occlusion in Doppler.
\section{Conclusion and Future Work}
\label{sec:conclusion_future_work}

This work introduced calibrated domain randomization as a simple but effective method for sim-to-real transfer in FMCW radar occupancy detection and people counting. By calibrating simulated range--Doppler noise floors to match global statistics of unlabeled real measurements, CDR significantly improves balanced accuracy on real test data compared to both baseline RT simulation-only training and uncalibrated domain randomization. The approach is lightweight, interpretable, and deployment constraints, making it practical for radar-based human sensing systems built from modern simulators. Future work will focus on enabling angle-of-arrival estimation for people counting and conducting studies on different activity recognition tasks.

{
    \small
    \bibliographystyle{ieeenat_fullname}
    \bibliography{main}
}

\end{document}